\documentclass[aps,twocolumn,groupedaddress,showpacs,amssymb,prb,floatfix,preprintnumbers]{revtex4}
\usepackage{graphicx,color}

\newcommand{\LaOFFeAs}       {${\rm La} {\rm Fe}  {\rm As} {\rm O}_{0.9} {\rm F}_{0.1}$}

\newcommand{\AsDef}             {${\rm La} {\rm Fe} {\rm As_{1-\delta}} {\rm O}_{0.9} {\rm F}_{0.1}  $}

\begin{document}

\thispagestyle{myheadings}

\title[Unusual disorder effects in \AsDef\ as revealed by NMR
spectroscopy]{Unusuals disorder effects in
superconducting \AsDef\ as revealed by NMR spectroscopy}

\author{F.\ Hammerath$^1$, S.-L.\ Drechsler$^1$, H.-J.\ Grafe$^1$, G.\ Lang$^1$, G.\ Fuchs$^1$,
G.\ Behr$^1$, I.\ Eremin$^2$, M. M.\ Korshunov$^{2,3}$\footnote{Present address: Department of Physics, University of Florida, Gainesville,Florida 32611, USA}, and B.\ B\"uchner$^1$}

\affiliation{$^1$IFW Dresden, Institute for Solid State Research, P.O. Box 270116, D-01171 Dresden, Germany\\
$^2$Max-Planck-Institut f\"ur Physik komplexer Systeme, D-01187 Dresden, Germany\\
$^3$L.V. Kirensky Institute of Physics, Siberian Branch of Russian Academy of Sciences, 660036 Krasnoyarsk, Russia}

\date{\today}

\begin{abstract}
We report $^{75}$As NMR measurements of the spin-lattice relaxation in the superconducting state of LaFeAsO$_{0.9}$F$_{0.1}$ and
As-deficient LaFeAs$_{1-\delta}$O$_{0.9}$F$_{0.1}$. The temperature behavior of $1/T_1$ below $T_c$ changes drastically from a $T^3$-dependence for LaFeAsO$_{0.9}$F$_{0.1}$ to a $T^5$-dependence for the As-deficient sample.
These results, together with the previously reported unexpected increase of $T_c$ and the slope of the upper critical field near $T_c$ for the As-deficient sample, are discussed in terms of non-universal SC gaps in Fe-pnictides
and the effect of As deficiency as an exotic case where nonmagnetic 'smart' impurities even stabilize an $s_{\pm}-$wave
superconductor or within a scenario of a disorder-driven change to $s_{++}$-superconductivity.
\end{abstract}


\maketitle

%%%%%%%%%%%%%%%%%%%%%%%%%%%%%%%%%%%%%%%%%%%%%%%%%%%%%%%%%%%%%%%%%%%%%%%%%%%%%%%%%%%%%%%%%%%%%%%%%%%%%%%%%%%%%%%%%%

The symmetry of the order parameter and the underlying Cooper-pairing mechanism
in the newly discovered Fe-based superconductors\cite{kamihara} are one of the most challenging problems in contemporary solid state physics. Historically, nuclear magnetic resonance (NMR) studies showing up the so called Hebel-Slichter peak
in the nuclear spin-lattice relaxation rate (NSLRR) played a significant role in establishing the BCS theory
as the first microscopic description of conventional (weakly coupled) superconductors.\cite{Slichter2008} Physically, this behavior is caused by the coherence factors and the symmetry of a single nodeless
superconducting (SC) gap.\cite{masuda} Nowadays, within a simplified approach
(ignoring damping, strong coupling, anisotropy, impurity, and
inhomogeneity effects \cite{Mitrovic06,Akis91}) its presence or absence
together with the $T$-dependence of the  NSLRR, $1/T_1$, below $T_c$ are frequently used to discriminate tentatively
conventional from unconventional pairing.
For a single Fermi surface (FS) sheet and superconductivity in the clean limit
$T^3$- and $T^5$-dependencies would be regarded as evidence for line- and point-node SC order
parameters, respectively, which for singlet pairing correspond to the $d$- and a
special $s+g$-wave state. Recently it has been realized that the situation in multibands
and especially in Fe-pnictides with impurities is far from
being that simple, in particular, there is no universal behavior for the growing number of related compounds.\cite{remark} 
The $s_{\pm}$-scenario proposed\cite{mazin,kuroki08,kuroki09,chubukov} at the early stages of the
Fe-pnictide research at present is still the most popular one. Due to the
vicinity of a competing spin density wave state in the phase diagram, it is tempting to assume that
antiferromagnetic (AFM) spin fluctuations might be the dominant pairing glue.
Then, from the FS topology given by small hole (electron) pockets centered around the
$\Gamma=(0,0)$ ($M=(\pi,\pi)$)-points of the Brillouin zone, a nodeless gap
with opposite signs on each of the disconnected FS pockets separated by the
wave vector ${\bf Q} = (\pi,\pi)$ is naturally suggested.

With respect to pair-breaking interband impurity scattering some doubts about this sign-reversed $s_{\pm}$-scenario have been put forward. \cite{Sato2009,kontani,kulic,Onari,Lee2009,Kontani09,Yanagi09}
Also the available weak coupling fits of the upper critical fields $B_{c2}(T)$ for the Nd-1111, the Sm-1111, and the La-1111 systems,\cite{Jaroszynski2008,Lee22009,Haindl2008}
all closely related to the ones considered here, do not show a dominant interband pairing interaction generic for the
intended $s_{\pm}$-scenario but result at most in comparable intraband and interband coupling
strengths or even in dominant intraband ones. Finally, for the two-gap system FeSe$_{1-x}$ only a tiny
interband coupling has been derived from the $T$-dependence of the penetration depth.\cite{khasanov2009}

Various experiments have been carried out to extract the symmetry of the SC order parameter
in Fe-based superconductors. In particular, ARPES and microwave data \cite{nodeless} are consistent with
a SC gap being nodeless on each FS pocket. These results taken together with the observation of
a peak at the AFM wave vector ${\bf Q}$ and $\omega=\omega_{res}$ found below $T_c$ in various
compounds\cite{res-peak} by means of inelastic neutron scattering (INS) experiments provide support in favor of
$s_{\pm}$-wave symmetry.  Note that a sharp resonance peak is a result
of different signs of the SC gap for {\bf k} and {\bf k+Q} points generic for the $s_{\pm}$-wave symmetry.
However, it has been argued recently \cite{Onari} that a somewhat broader peak-like feature can be attributed to a self-energy renormalization of quasiparticles in $s_{++}$-wave (sign preserved) superconductors. A similar feature in Raman spectra has not been observed.\cite{Hackl2009} Hence, the assignment of the observed INS features is controversial. 

In this unclear situation we report $^{75}$As NMR measurements of the NSLRR in LaFeAsO$_{0.9}$F$_{0.1}$ and As-deficient LaFeAs$_{1-\delta}$O$_{0.9}$F$_{0.1}$.
Surprisingly we observe a drastic change of the $T_1^{-1}(T)$ dependence below $T_c$ from $T^3$ for LaFeAsO$_{0.9}$F$_{0.1}$ to $T^5$ for LaFeAs$_{1-\delta}$O$_{0.9}$F$_{0.1}$. 
Comparing our NMR data with other available data, we discuss three alternative scenarios: (i) a
non-universal superconducting gap, (ii) a disorder driven transition from $s_{\pm}$-
to $s_{++}$-wave symmetry in Fe pnictides as well as (iii) the As deficiency as
a rare case of defects which can yield even a stabilization of unconventional $s_{\pm}-$wave
or conventional $s_{++}$-multiband superconductivity. Theoretical issues to be settled in
(ii) and (iii) as well as further experiments to clarify the challenging situation are proposed.

\begin{figure}[t]
\begin{center}
\includegraphics[width=\columnwidth,clip]{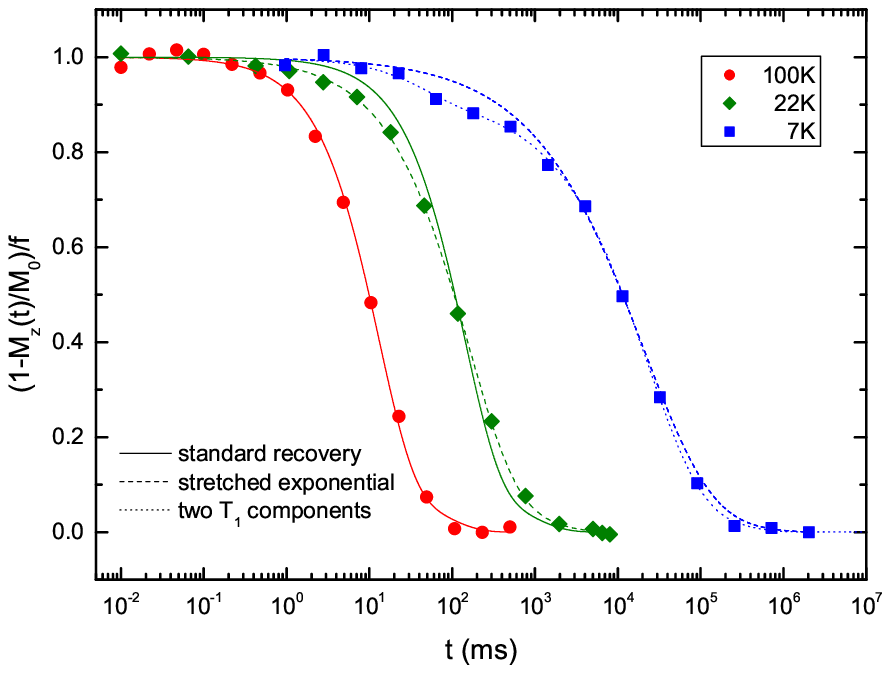}
 \caption{(color online) Recovery curves for $T=100$ K (red points), $T=22$ K (green diamonds)
 and $T=7$ K (blue squares) in $H_0 = 7.01$ T. Normalization corresponds to a division by the prefactor $f$(=1.7-2) of Eq. (\ref{singleexponential}). 
The lines are examples for the different fitting functions containing a single $T_1$ component (solid line), a distribution around one $T_1$ component with a stretching parameter $\lambda$ (dashed line) and two components $T_{1sc}$ and $T_{1s}$ (dotted line).}
\label{recovery}
\end{center}
\end{figure}

A polycrystalline sample of \AsDef\ was prepared by standard
methods and characterized by x-ray diffraction, susceptibility and resistivity measurements.\cite{Fuchs2008,Fuchs2009}
The As-deficiency was obtained by wrapping the sample in a Ta foil during the annealing procedure
leading to $\delta$=0.05-0.1. The increased disorder is reflected in the enhanced resistivity in the normal state compared to the clean sample.\cite{Kondrat} However, $T_c$ and the slope of $B_{c2}(T)$ near $T_c$ {\it increase} unexpectedly from 26.8~K and -2.5~T/K in the stoichiometric compound to 28.5~K and -5.4 T/K in the As-deficient compound.
$\mu$SR measurements proved an enhanced paramagnetism, which is the origin of the observed Pauli-limiting behavior of $B_{c2}(T)$ at lower temperatures.\cite{Fuchs2009}
The SC volume fraction is about 90\% while in the pure sample it amounts to 100\%. \cite{LuetkensNatMat}

For the NMR experiments the sample was ground to a powder.
The $^{75}$As NMR spectrum showed a typical powder pattern as reported previously.\cite{Grafe2008}
The $^{75}$As NSLRR $T_1^{-1}$ was measured at the peak corresponding to $H\| ab$
in a magnetic fields of $H_0$ = 7.01 T using inversion recovery.
The recovery of the longitudinal magnetization $M_z(t)$ was fitted to
the standard expression for magnetic relaxation of a nuclear spin
of $I=3/2$ which reads:
\begin{equation}
M_z(t)=M_0[1-f(0.9\rm{e}^{-(6t/T_1)^{\lambda}}+0.1\rm{e}^{-(t/T_1)^{\lambda}})] \label{singleexponential}
\end{equation}

\begin{figure}[t]
\begin{center}
\includegraphics[width=\columnwidth,clip]{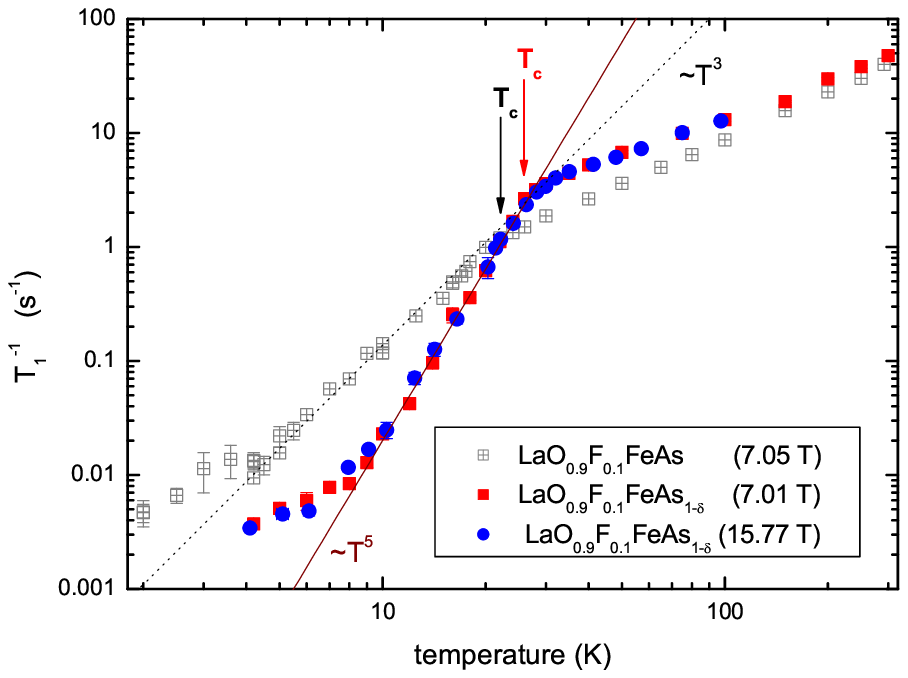}
\caption{(color online)$^{75}$As SLRR  for \AsDef\ in 7.01 T (red squares) and 15.77 T (blue circles) compared to \LaOFFeAs\ in 7.05 T (grey crossed squares \cite{Grafe2008}, new data points for $T\leq 4.2$ K). The dotted line illustrates the $T^3$ behaviour of $T_1^{-1}$ for \LaOFFeAs, the solid line indicates the $T^5$ behaviour observed for \AsDef.}
\label{T1}
\end{center}
\end{figure}

Typical recovery curves for $T=100$~K, $T=22$ K and $T=7$ K are given in
Fig. \ref{recovery}. Above $T_c(H_0) \approx 26$ K the recovery
could be nicely fitted with a single $T_1$ component ($\lambda =1$).
For $T<T_c$ a stretching parameter
$\lambda < 1$ was needed to account for a distribution of
NSLR times around a characteristic relaxation time.
For $T\leq14$ K, where the intrinsic
NSLR time  $T_{1sc}$ in the SC state amounts to a few seconds, we
could distinguish a second, short contribution $T_{1s}$.
For this $T$-range a fitting function containing 
two weighted $T_1$ components was used. While the determination
of $T_{1s}$ was imprecise,
the long time component $T_{1sc}$, which displays the intrinsic
relaxation in the SC state, did not depend on the fitting procedure. 
$T_{1s}$ lies in the range of several hundred ms, indicating non-SC regions in the sample. Its weight of $(20 \pm 10)$ \% suggests, in addition to vortex cores, a minority non-SC volume fraction in agreement with $\mu$SR-measurements.
Fig. \ref{T1} shows the $T$-dependence of the $^{75}$As NSLRR
$T_1^{-1}$ for the As-deficient sample \AsDef\ and that of a
sample with the same F content, but without As-vacancies.\cite{Grafe2008} For the latter one, recently measured additional data points for $T\leq 4.2$ K are shown. Very surprisingly, for $T<T_c$ the
NSLRR of \AsDef\ decays with $T^5$, in contrast to the $T^3$-dependence 
of \LaOFFeAs. Using a field of 15.77 T this unexpected behavior was preserved within error bars, as shown in fig. \ref{T1}.

For $T \leq 8$~K, $T_1^{-1}$ of \AsDef\ deviates from this $T^5$ behavior and
changes to a linear $T$-dependence. A similar behavior is also visible for \LaOFFeAs\ for $T \leq 4.2$~K.
These low-$T$ features with a nearly linear slope below $T \approx 0.3 T_c$ were also observed in  BaFe$_2$(As$_{0.67}$P$_{0.33}$)$_2$  and explained with the existence of a residual density of states (RDOS) in a line-node model.\cite{nakai_new} Such a high RDOS can be excluded by penetration depth data derived from $\mu$SR for the La1111 samples \cite{LuetkensNatMat} as well as for other pnictide systems.\cite{nodeless} 
Among other possible mechanisms, the classical spin diffusion \cite{Saito07} is unlikely due to the lack of field dependence of $T_1^{-1}$. Another possibility are thermal fluctuations of vortices, which induce alternating magnetic fields contributing to the relaxation.\cite{Ehrenfreund}
Further study on a larger field range, especially at low fields, is needed to clarify this $T_1^{-1} \propto T$ behavior.

We will now discuss the different $T$-dependencies for $ T > 0.3 T_c$ within the previously mentioned scenarios.
To the best of our knowledge, no exponential but power-law dependencies $\propto T^n $ have been observed \cite{Grafe2008,mukuda,Terasaki,Nakai08,Fukazawa,zheng,Kawasaki,Kobayashi2009,Sato2009,zheng2,yashima,ning,Zhang}
for all Fe-based pnictide superconductors, with $n$ in between 1.5 and 6, indicating unconventional superconductivity. These power law dependencies have been discussed within different models, such as $s_{\pm}$- and d-wave symmetries, including the possibility of multiple SC gaps.\cite{zheng,Kawasaki,zheng2,yashima}
Within the 122 family heavily overdoped compounds such as KFe$_2$As$_2$
exhibit the lowest value observed so far whereas optimally or slightly underdoped compounds show the largest values.
Recently, it has been suggested,\cite{Sato2009,Kobayashi2009} that the frequently observed $T^3$ power-law in the NSLRR should not be considered as an intrinsic effect but instead be attributed to some unspecified inhomogeneities in view of the missing correlation between the $T_c$-value and the NSLRR exponent, while higher exponents would occur for cleaner samples. However, we find just the opposite behavior. In Fig.\ 3 we show the normalized $T_1^{-1}(T)$ curves for our As-deficient sample and the samples from Ref. \onlinecite{Sato2009}. Their nominal clean sample as well as the Co-doped one exhibit
nearly the same $T_1^{-1}(T)$-dependence as our As-deficient sample, whereas our clean sample exhibits the $T^3$-dependence (see Fig \ref{T1}). In our opinion this points towards sizeable disorder in the samples of Ref.\ \onlinecite{Sato2009}. This is further supported by the lowest resistivity of our clean sample compared to all others.\cite{Fuchs2009, Sato2009} In this context a measurement of the upper critical field on the same samples would be helpful to clarify this point and to elucidate the role of impurities/vacancies in Fe-based superconductors in general.

In principle, our observation of an unusual transition from $T^3$ to $T^5$ with {\it increasing} disorder is not necessarily inconsistent with a $s_{\pm}$-wave SC gap though alternative scenarios should be invoked, too. Starting from the clean limit it has been shown \cite{scalapino,kuroki09,chubukov_nodal} that within the generalized $s_{\pm}$-wave scenario both node-less and nodal SC gaps might occur depending on the proximity of the doped sample to the AFM
instability. In this regard, naively our finding can be interpreted in favor of a transition from the nodal to the
nodeless SC gap upon adding As defects which for some reason might drive the system closer to
antiferromagnetism, in accord with the slightly enhanced normal state NSLRR and the slightly changed lattice constants \cite{Fuchs2009} of the As deficient sample.

However, such a simplistic point of view cannot be easily applied to pnictides as it is also known that the
$s_{\pm}$-wave ground state is sensitive to non-magnetic impurities. Most importantly,
the intraband impurity scattering does not affect the superconductivity, since the SC gap does not change its
sign within each of the bands. At the same time the scattering with large momenta which connects electron and hole pockets (interband scattering) is pair-weakening and thus yields a decrease of $T_c$ and simultaneously introduces power laws in the thermodynamics and $1/T_1$ at intermediate temperatures.
Therefore, if for some reason As vacancies act as  'smart' impurities which change the ratio between the intra- and
interband scattering, our observations could be also explained. These changes have to
be reflected similarly also in the other thermodynamical or transport properties such as
penetration depth or thermal conductivity. Unfortunately, there is no direct way to estimate the ratio of the
intraband to interband scattering rates from experiments since usual characteristics like the residual resistance ratio or the mean free path are quantities which mostly indicate the overall impurity effects but not their ratio.
\begin{figure}[t]
\begin{center}
\includegraphics[width=\columnwidth,clip]{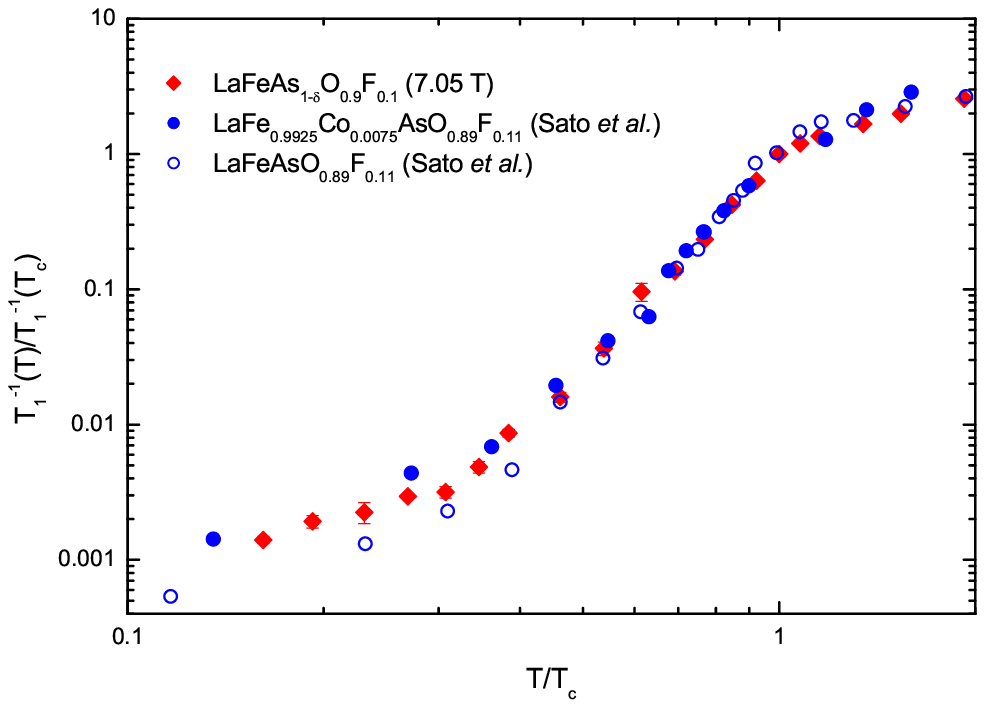}
\caption{(color online) Comparison of the $^{75}$As NSLRR for our \AsDef\ (red diamonds) with pure and Co-doped ${\rm La} {\rm Fe}  {\rm As} {\rm O}_{0.89} {\rm F}_{0.11}$ from Ref.\ \onlinecite{Sato2009} (open and filled circles).}
\label{Sato}
\end{center}
\end{figure}

Thus, the above-mentioned scenario is based on the assumption that
$s_{\pm}$-wave order is stable and adding As vacancies either changes the proximity to the
competing antiferromagnetism or/and the ratio of intra- to interband non-magnetic impurity scattering in
pnictides. There is, however, another intriguing possibility.
Let us assume that there is a substantial electron-boson interaction
which provides an attractive intraband potential for Cooper-pairing.
In this case a (weak) repulsive interband Coulomb scattering
will still lead to the $s_{\pm}$-wave SC order in the clean limit though the
attractive electron-boson interaction dominates. However, once the As vacancies change the ratio between intra- and interband impurity scattering, a transition from $s_{\pm}$-wave to conventional $s_{++}$-wave SC order may be induced.
This scenario, however, still needs further experimental clarification. For example, despite
the transition from $T^3$ to $T^5$ behavior we do not find any sign of the
Hebel-Slichter peak in the latter case close to $T_c$. Moreover, current
experimental data on the importance of the electron-phonon coupling
are not very conclusive. Therefore, the intriguing possibility of high-energy charge
fluctuations as well as weak electron-phonon interactions with orbital fluctuations \cite{Kontani09,Yanagi09}
deserve more detailed studies. Another interesting point would be a detailed comparison
with FeSe$_{0.92}$, which exhibits a $T^3$-law for $1/T_1$ (Ref. \onlinecite{Kotegawa2008}) and other Fe-based
superconductors with vacancies in the polarizable subsystem.

To summarize, we present
challenging NMR-experimental data on disordered As-deficient
samples. We strongly believe that a future quantitative realistic theoretical description of our data within unconventional $s_{\pm}$- or conventional, but unusual $s_{++}$-superconductivity scenarios will stimulate the further development
of these approaches and this way be finally helpful for the elucidation of the underlying but
yet unsettled mechanism.

%%%%%%%%%%%%%%%%%%%%%%%%%%%%%%%%%%%%%%%%%%%%%%%%%%%%%%%%%%%%%%%%%%%%%%%%%%%%%%%%%%%%%%%%%%%%%%%%%%%%%%%%%%%%%%%%%%%%%%%%%%%%%%%%%%%%%%%%%%%

We thank M. Deutschmann, S.\ M\"uller-Litvanyi, R.\ M\"uller, R.\
Vogel, and A.\ K\"ohler for experimental support. This work has
been supported by the DFG, through FOR 538, SPP 1458 and Contract No.\
Be1749/12. GL acknowledges support from the A.\ v.\
Humboldt-Stiftung. MMK acknowledges support from RFBR 09-02-00127, RAS program on ``Low temperature quantum phenomena'' and President of Russia MK-1683.2010.2. SLD thanks H.\ Kontani, K.\ Kuroki, K.\ Kikoin and M.\ Kuli{\'c} for useful discussions. \\


\begin{thebibliography}{99}
\bibitem{remark} Note that early penetration depth studies on
LaFePO (T$_c \sim$ 6K) suggested nodal superconductivity,\cite{carrington} which is inconsistent with
a fully-gapped $s_{\pm}$-wave gap.\cite{mazin,chubukov,kuroki08} Furthermore, recent low-$T$ studies of the thermal conductivity in BaFe$_2$(As$_{1-x}$P$_x$)$_2$ revealed a nodal structure of the SC gap. These findings taken together with other data which are consistent with a nodeless structure point to the non-universality of the SC Fe-based materials.
\end{thebibliography}
\end{document}